\documentclass[a4paper,UKenglish,cleveref,autoref]{lipics-v2021}

\pdfoutput=1 
\hideLIPIcs  

\nolinenumbers

\usepackage{booktabs}

\title{Sequencelib: A Computational Platform for Formalizing the OEIS in Lean}

\author{Walter {Moreira}}{Texas Advanced Computing Center, University of Texas at Austin, Austin TX USA}{walter@waltermoreira.net}{https://orcid.org/0000-0003-4830-687X}{}{}

\author{Joe Stubbs}{Texas Advanced Computing Center, University of Texas at Austin, Austin TX USA}{jstubbs@tacc.utexas.edu}{https://orcid.org/0000-0002-8644-0300}{NSF awards \#1931439 and \#2112606}

\authorrunning{W. Moreira and J. Stubbs} 

\Copyright{Walter Moreira and Joe Stubbs} 



\category{} 

\relatedversion{} 

\acknowledgements{The authors would like to thank Vijay Ganesh and his students Leyan, Greg, Prithwish, and Congyan (Cruise), for many helpful discussions as part of an ongoing collaboration related to Sequencelib.}

\EventEditors{John Q. Open and Joan R. Access}
\EventNoEds{2}
\EventLongTitle{42nd Conference on Very Important Topics (CVIT 2016)}
\EventShortTitle{CVIT 2016}
\EventAcronym{CVIT}
\EventYear{2016}
\EventDate{December 24--27, 2016}
\EventLocation{Little Whinging, United Kingdom}
\EventLogo{}
\SeriesVolume{42}
\ArticleNo{23}

\lstdefinelanguage{lean}{
    literate=
        *{ℕ}{{\ensuremath{\mathbb{N}}}}{1}
        {→}{{\ensuremath{\rightarrow}}}{1}
        {∀}{{\ensuremath{\forall}}}{1}
}
\begin{document}

\maketitle

\begin{abstract}
The On-Line Encyclopedia of Integer Sequences (OEIS) is a web-accessible database cataloging interesting integer sequences and associated theorems. With more than 12,000 citations, the OEIS is one of the most highly cited resources in all of theoretical mathematics. In this paper, we present Sequencelib, a project to formalize the mathematics contained within the OEIS using the Lean programming language. Sequencelib includes a library of Lean formalizations of OEIS sequences as well as metaprogramming tools for programmatically attaching OEIS metadata to Lean definitions and deriving theorems about their values. Further, we describe OEIS-LT, a highly scalable Lean server that exposes these tools via a low-latency API. Finally, using OEIS-LT and prior work of Gauthier, et al. \cite{gauthier2023learning}, we describe a computational pipeline that formalized more than 25,000 sequences from the OEIS and proved more than 1.6 million theorems about their values. Our method makes use of a transpiler, available in OEIS-LT, that is capable of translating a subset of Standard ML to Lean, together with a set of performance improvement transformations and proofs of correctness. 
\end{abstract}

\section{Introduction}

The On-Line Encyclopedia of Integer Sequences (OEIS) is a web-accessible catalog and
search facility for discovering integer sequences. With more than 12,000
citations, the OEIS is one of the most highly cited resources in all of theoretical
mathematics. The OEIS contains more than 385,000 sequence definitions, each described
in natural language, together with a wealth of metadata about the sequences, including
many known values, propositions, conjectures, and relations between them.

In this paper, we present Sequencelib, a computational platform to formalize the
mathematics contained within the OEIS using the Lean 4 programming language. Formalization
of the OEIS would: 1) remove ambiguities from natural language descriptions,
both within the sequence definitions themselves and within notes and other
metadata describing properties of the sequences; 2) ensure the encyclopedia is
free of errors or other inaccuracies; 3) produce formalizations of a rich set of
mathematics across a range of areas; 4) serve as a benchmark for evaluating
theorem provers and AI tools; and 5) provide a programmable interface to study
the OEIS as a whole, including dependencies between sequences.

The Sequencelib platform provides a computational ecosystem that supports both human
and automated formalization, and a central goal of the project is to facilitate
code generation and proof synthesis with AI-based tools. The primary components of
Sequencelib are: 1) a library of Lean code containing formalizations of sequence
definitions from OEIS and theorems about them; 2) a new Lean tactic and a metaprogramming
attribute as well as associated functions for automatically attaching OEIS metadata
to Lean function definitions and deriving theorems about their values; 3) an automated
build and documentation generator that programmatically collects all sequences and
theorems contained within the code library and generates a high-quality,
searchable website; and 4) OEIS-LT, a multi-threaded, highly scalable tool
server, written in Lean, that can be used by both humans and automated systems
to interact with the Sequencelib platform.

In subsequent sections, we provide details on the design and implementation of the
primary components of the Sequencelib ecosystem. Furthermore, we describe a computational
pipeline that leveraged Sequencelib to formalize more than 25,000 OEIS sequence definitions
and to prove more than 1.6 million theorems about their values. This pipeline
builds on work of Gauthier, et al. \cite{gauthier2023learning} and leverages a
transpiler within OEIS-LT that can translate a subset of ML to Lean 4. The pipeline
establishes the scalability and effectiveness of the Sequencelib tools.

The rest of the paper is organized as follows: Section \ref{sec:background}
collects necessary background material and related work; Section \ref{sec:sl_overview}
provides an overview of Sequencelib and the goals for the project; in Section
\ref{sec:sl_meta} we describe the design and implementation of Sequencelib's
metaprogramming functions; Section \ref{sec:pipeline} describes the OEIS-LT server
and the autoformalization pipeline; in Section \ref{sec:future} we provide an
overview of our future plans for the platform before concluding in Section
\ref{sec:conclusion}.

\section{Background and Related Work}
\label{sec:background}
In this section we review background material necessary for the rest of the manuscript. 

\subsection{Interactive Theorem Provers and Lean}
Interactive Theorem Provers (ITPs), such as Lean \cite{moura2021lean}, Isabelle \cite{nipkow2014concrete}, and Rocq (formerly named Coq) \cite{huet1997coq}, are tools for formalization that combine a programming language that is capable of expressing mathematical statements with tools that enable humans to guide the proof creation process. Based on the Calculus of Inductive Constructions, Lean is capable of expressing theoretical mathematical statements and their proofs. Additionally, Lean can be viewed as a programming language with robust metaprogramming facilities that allow for the automation and extension of Lean itself.

Powerful advances in ITPs have led to their increased use in advanced cases of research mathematics, such as the formalization of Polynomial Freiman-Ruzsa (PFR) Conjecture\cite{poly_fr_conj}, Perfectoid spaces \cite{buzzard2020formalising}, optimal sphere packing (the $E_8$ lattice) \cite{sphere_packing_lean} and Fermat's Last Theorem \cite{fmt_lean}. Prominent mathematicians, including a number of Fields Medal winners, such as Peter Sholze, Terrance Tao, and Maryna Viazovska, have strongly advocated for their use. Moreover, a growing number of university math departments are exploring complementing traditional undergraduate curricula with ITP-based content. 

\subsection{Mathlib} Mathlib \cite{mathlib}, written in the Lean programming language, 
is one of the fastest growing libraries of formal mathematics. At nearly 2 million lines of code, Mathlib contains a significant portion of advanced mathematics across all major branches, including algebra, analysis, geometry, number theory, and topology. The depth and quality of mathlib qualify it for use in the formalization of research mathematics, and it has been used to that end in many projects, see, e.g., \cite{mathlib_research_1, mathlib_research_2, mathlib_research_3, mathlib_research_4}.

The origins of the Sequencelib project began with an effort to contribute to Mathlib a formalization of some results about Sylvester's sequence, that appeared as an exercise in the textbook ``Concrete Mathematics''~\cite{knuth1989concrete}. After a discussion with some of the primary Mathlib maintainers, it was decided that the results on Sylvester's sequence would not be appropriate for inclusion in Mathlib and would better suited in a separate library dedicated to results on mathematical sequences. In response, we launched the Sequencelib project to focus on formalizations of integer sequences included in the On-Line Encyclopedia of Integer Sequences (OEIS) with the intention of leveraging and contributing to Mathlib when appropriate. 

\subsection{The On-Line Encyclopedia of Integer Sequences}

The On-line Encyclopedia of Integer Sequences (OEIS) \cite{oeis_web} is a database of integer sequences, defined in natural language, together with known values of each sequence and additional metadata. The OEIS was started by Neil Sloane in 1964 and transitioned to the OEIS Foundation in 2009. It contains more than 390,000 sequences and has maintained a consistent growth of approximately 10,000 new sequences each year \cite{oeis_faq}. The OEIS has had a profound impact on mathematical research, with more than 12,000 citations as of October, 2025 \cite{oeis_citations}.

\subsection{Proof Synthesis, Autoformalization, and Tool Servers}
Despite the growing advances and popularity of projects like Lean, the use of ITPs to formalize mathematics is still notoriously difficult, requiring substantial expertise beyond the specific mathematical subject area. As a result, \textit{proof synthesis}, i.e., the task of constructing a proof of a formal mathematical statement in an automated manner, has become an important challenge. Quite recently, an exciting new direction involving the integration of formal methods and AI has emerged. AI-enabled systems have been used on a variety of formalization tasks, including \textit{auto-formalization}, that is, encoding natural language descriptions of theoretical mathematics into formal descriptions, such as a Lean program, and \textit{auto-informalization}, i.e., providing natural language descriptions of formalized mathematics, have found surprising success. In many cases, successful approaches combine AI models with domain-specific \textit{tools servers} which provide programmable interfaces to specialized methods. For example, Google’s AlphaProof \cite{deepmind2024alphaproof} and AlphaGeometry~\cite{trinh2024solving} combined deep learning and reinforcement learning with tools developed in Lean to obtain a silver medal at the International Mathematics Olympiad \cite{trinh2024solving}. 
Moreover, there is growing interest in using formal methods to improve the robustness and reliability of AI/ML, particularly large language models, which are notoriously susceptible to hallucinations. 

\section{Overview of Sequencelib}
\label{sec:sl_overview}

The goal of the Sequencelib project is to provide a platform for formalizing the mathematics contained within the OEIS in Lean. In this context, formalization of an OEIS sequence includes three components: 1) one or more Lean functions, $f : \mathbb{N} \to \mathbb{N}$ or $f : \mathbb{N} \to \mathbb{Z}$, with domain $\mathbb{N}$ and co-domain either $\mathbb{N}$ for non-negative sequences, or $\mathbb{Z}$ for signed sequences, whose values coincide with the values of the sequence;
 2) theorems of the form $f(n) = b$ for known values $b$ of the sequence; and 3) theorems of the form $\forall n \in\mathbb{N}, f(n) = g(n)$ whenever there are multiple definitions of the same sequence. Sequencelib leverages Mathlib whenever applicable, but it also includes definitions and theorems related to sequences that would be out of scope for Mathlib. In particular, Sequencelib welcomes formalizations of any concepts or theorems relevant to integer sequences even if they are not in the most general or abstract form. Moreover, Sequencelib values computable definitions of sequences even when non-computable definitions may be readily available in Mathlib.

For example, for sequence \textsc{A000045}, the famous Fibonacci numbers, Sequencelib provides two definitions. The first leverages the definition from Mathlib: \texttt{Sequence.Fibonacci = Nat.fib}. The second definition is shown in the following listing and closely resembles an implementation in an imperative programming language.

\begin{lstlisting}[caption={A second definition of the Fibonacci sequence}]
def Fibonacci2 (n : ℕ) : ℕ := Id.run do
  let mut x := 0
  let mut x_prev := 0
  let mut y := 1
  for _ in [0:n] do
    x_prev := x
    x := y
    y := x_prev + y
  pure x
\end{lstlisting}

In this case, both functions are \texttt{computable} in the Lean sense, but in other cases, one or more definitions may be (Lean) \texttt{noncomputable}. For example, Sequencelib includes the following definition of the powers of primes sequence, A000961, based on the \texttt{IsPrimePow} predicate from Mathlib, as well as an alternative definition that is computable. 

\begin{lstlisting}[caption={A noncomputable definition of the powers of primes}]
noncomputable def PowersOfPrimes : ℕ → ℕ
  | 0 => 1 
  | 1 => 1
  | n + 1 => nth IsPrimePow (n - 1)
\end{lstlisting}

Deriving theorems about the values of (Lean) \texttt{noncomputable} definitions can be more challenging than proving values about computable functions. This point is discussed in more detail in Section \ref{sec:sl_meta}.

The OEIS contains sequences whose definitions and values involve deep and advanced theoretical mathematics, including many open problems and conjectures. As a result, Sequencelib's goal of formalizing the entire OEIS presents a formidable challenge. For example, sequence A000001 from the OEIS counts the number of non-isomorphic finite groups. Counting such groups involves the classification of finite simple groups (the so-called enormous theorem), whose proof consists of thousands of pages of mathematics developed over 60 years by more than 100 mathematics in the mid-20th and early 21st century. Sequencelib includes two definitions for A000001. First, a category-theoretic definition using objects in the category \texttt{Grp} from Mathlib is given. Then, a sequence counting the non-isomorphic subgroups of order $n$ of $S_n$, the symmetric group on \texttt{n} letters, is defined. These two definitions are equivalent, and Sequencelib includes a proof based on Cayley's theorem, which has been formalized in Mathlib. 


The Sequencelib Lean 4 library is hosted in a public repository on Github \cite{sequencelib_github}, and the project welcomes contributions (see the official contributing guidelines on the website). From the Lean code we run a continuous integration pipeline, on each commit to the main branch, which performs several tasks: (1) it builds the library, (2) it runs an external checker (\texttt{lean4checker}\footnote{\url{https://github.com/leanprover/lean4checker}}), and (3) it generates a website containing information about all formalized sequence definitions and theorems. Sequencelib includes a Lean attribute, \texttt{OEIS}, that allows Lean function definition to be programmatically linked to an OEIS sequence, as shown in the following listing.

\begin{lstlisting}[caption=An example of tagging a function with the OEIS attribute]
@[OEIS := A000079]
def Powerset (n : ℕ) : ℕ := 
  (Finset.range n).powerset.card
\end{lstlisting}

The Sequencelib website generator traverses the Sequencelib abstract syntax tree and collects definitions and theorems. Moreover, the \texttt{OEIS} attribute provides a hook into Sequencelib's \texttt{oeis-tactic}, which can automatically prove theorems about the values of sequences in certain cases. More details about Sequencelib's metaprogramming and the \texttt{oeis-tactic} is given in Section \ref{sec:sl_meta}.

In addition to human collaboration, a primary goal for the project is to enable autoformalization and proof synthesis via AI-enhanced systems. The Sequencelib platform provides a scalable tool server, OEIS-LT, written in Lean, that presents a web-friendly JSON API over a Unix domain or TCP socket. A high-performance deployment of the server available for public use is hosted at the Texas Advanced Computing Center at the University of Texas at Austin. Additionally, the tool server may be utilized via a high-quality Python SDK. More details about OEIS-LT are provided in Section \ref{sec:pipeline}.


\section{Sequencelib Metaprogramming Tools and Tactics}
\label{sec:sl_meta}

The main goal for Sequencelib as a platform for formalizing integer sequences is to provide an environment that facilitates the work for humans and AI systems. To fulfill part of that objective, we provide a set of tools that allow for automatic indexing and documentation of sequences, as well as automatic generation of theorems for their values. These tools rely on the metaprogramming capabilities of Lean that enable programmatic inspection of the source code and the elaborated values (i.e., the values that Lean's kernel verifies). Metaprogramming in Lean also gives the ability of constructing new terms and type checking them through the kernel.

The entry point for Sequencelib metaprogramming is the \texttt{OEIS} attribute. Sequencelib processes and adds the definition of a function that uses this attribute to an internal table, which is then used for a variety of tasks such as generating documentation or theorems. The following is an example of a definition using the \texttt{OEIS} attribute:

\begin{lstlisting}[caption={An example of tagging a function with the OEIS attribute}]
@[OEIS := A000079, offset := 0, maxIndex := 10, derive := true]
def PowersOfTwo (n : ℕ) : ℕ := 2 ^ n
\end{lstlisting}

In this example, we declare that the definition of the function \texttt{PowersOfTwo} represents the OEIS sequence with tag A000079. The \texttt{offset} indicates the initial index for the sequence, as given by the OEIS entry. Values of the function for indices smaller than the offset are ignored from the point of view of the documentation and the generation of theorems. The attributes \texttt{maxIndex} and \texttt{derive} ask Sequencelib to generate theorems of the form:

\begin{lstlisting}[mathescape]
theorem thm_$i$ : PowersOfTwo $i$ = $j$
\end{lstlisting}

Index $i$ varies in $i=\texttt{offset},\ldots,\texttt{maxIndex}$, and Sequencelib computes the value $j$ using the Lean compiler. We restrict the generation of theorems for Lean-computable functions, hence making the computation of $j$ feasible. We have plans to relax this restriction in the future, by providing a way to obtain the values of the sequences without evaluating the Lean definitions, likely leveraging the OEIS-LT server. Generating the theorems happens right after Lean compiles the definition of the function. The \texttt{OEIS} attribute succeeds only when all theorems up to \texttt{maxIndex} are successfully proved and added to the environment, which guarantees that the kernel accepts them. As mentioned in Section~\ref{sec:sl_overview}, the continuous integration pipeline runs an external checker on the compiled code, to ensure that the metaprogramming code is not introducing unsoundness.

Sequencelib attempts to prove the theorems that it auto-generates by using a new tactic: \texttt{oeis\_tactic}. This tactic includes a collection of tactics from Mathlib that were experimentally found to be successful for the particular form of theorems considered. In particular, it is fine tuned to handle the definitions generated from the pipeline described in Section~\ref{sec:pipeline}. In its current form, this tactic does not perform complex strategies, choosing instead to delegate the work to the selected tactics from Mathlib. However, consolidating the curated list of tactics in a single instruction allows for a modular approach. We have plans to extend the \texttt{oeis\_tactic} to handle additional cases, so that Sequencelib can auto-derive theorems from other machine learning pipelines. 

In addition to generating theorems, the \texttt{OEIS} attribute also collects an internal table of information for the sequences. That information is used to augment the documentation that Lean generates, for building an indexed website, and for providing input to the OEIS-LT server. The latter is an important aspect of Sequencelib since it enables programmatic access to the data, either by humans or by AI agents. The internal table contains data collected from the parameters of the \texttt{OEIS} attribute (such as the tag, the \texttt{offset}, and the generated theorems), and it also contains data computed from the environment.

The data that Sequencelib computes for a sequence includes: Lean-computability (whether the definition contains the attribute \texttt{noncomputable}) and theorems about values and equivalences found in the environment. Sequencelib crawls its environment searching for theorems of the form \lstinline{f i = j}, and of the form \lstinline{∀ (n : ℕ), f n = g n}. The first form captures all the auto-generated theorems plus any other theorems for values provided by the user. The second form captures equivalences for sequences that have more than one definition. This information allows Sequencelib to classify the functions in classes of equivalences and display the information in the website.  In addition to populating the documentation, we attach some of this information to the definitions of the sequences. For example, it is possible to query the offset and the tag for a given definition by evaluating the attributes \texttt{OEIS} and \texttt{offset}, such as:

\begin{lstlisting}[mathescape]
#eval PowersOfTwo.OEIS    /- A000079 -/
#eval PowersOfTwo.offset  /- 0 -/
\end{lstlisting}

Attaching these attributes to the function makes the sequence self-contained by carrying the necessary information that cannot be derived from the formal expression, together with its definition.
Users can also interactively display the internal table with the full information of the environment by executing the commands:

\begin{lstlisting}[mathescape]
/- display table in the Lean InfoView -/
#oeis_info 
/- display table as a JSON object -/
#oeis_info_json
\end{lstlisting}

For more robust access to this table, either by humans developing a tool that uses Sequencelib, or by AI agents, we describe the OEIS-LT server in the next section.

\section{OEIS-LT and Autoformalization of the OEIS}
\label{sec:pipeline}

In this section, we provide an overview of OEIS-LT, a Lean tool server that supports the formalization of OEIS sequence definitions and theorems by humans and automated processes, such as AI agents. We also present an automated pipeline that leveraged OEIS-LT to formalize more than 25,000 OEIS sequence definitions and prove more than 1.6 million theorems. 

\subsection{Overview of OEIS-LT}
OEIS-LT is a lightweight, multi-threaded server, written in Lean, that supports multiple clients making simultaneous requests over a TCP or local Unix Domain socket. Client requests and OEIS-LT server responses are formatted using JSON, and OEIS-LT provides its primary functionality through the support of different \textit{commands}. At present, four primary commands are supported: 
\begin{description}
    \item[\texttt{gen}] -- Generate a function definition, in Lean, from a function definition in Standard ML.
    \item[\texttt{compile}] -- Attempts to compile a Lean source code.
    \item[\texttt{eval}] -- Evaluate a Lean function definition on a list of input integers, and compare the values to a list of input values. 
    \item[\texttt{prove}] -- For an input Lean function definition, $f$, and an input list of integers, $[(a_i, b_i)]$, attempts to prove a set of theorems of the form $f(a_i) = b_i$. \textit{Note:} the $b_i$ are optional, and if not provided, OEIS-LT will first to try to compute $f(a_i)$ before trying to prove the theorem. 
\end{description}
Additionally, the \texttt{ready} command can be used to check the status of the server. 

For example, to evaluate the Lean function \texttt{PowersOfTwo} on the inputs $0, 1, 2, 3$ and compare against the expected values \texttt{1, 2, 4, 8}, a client sends a request to the OEIS-LT with a JSON message formatted as in the following listing: 

\begin{lstlisting}[caption={An example of a JSON request to OEIS-LT to evaluate PowersOfTwo}]
{ 
  "cmd": "eval", 
  "args": {
      "src": "def PowersOfTwo (n : ℕ) : ℕ := 2^n",
      "values": [(0, 1), (1, 2), (2, 4), (3, 8)], 
      "tag": "A000079"
  }
}
\end{lstlisting}

Upon receiving the request, OEIS-LT first parses the \texttt{src} string into a Lean \texttt{syntax} object. It then checks that the syntax contains an actual definition and elaborates the syntax, ensuring that it type checks.
Finally, for each pair passed via the \texttt{values} parameter, OEIS-LT generates a syntax expression for function application, elaborates the syntax, and evaluates the Lean expression. Note that this evaluation assumes the function is \texttt{computable} in the Lean sense.

Similarly, for the \texttt{prove} command, OEIS-LT requires a \texttt{src} argument representing a Lean function, but it also requires a \texttt{values : Array (Nat × Option Int)} parameter specifying the input(s) to the function and, optionally, the target value(s) for the function at each input. As before, OEIS-LT proceeds to parse the \texttt{src} string into a Lean \texttt{syntax} object, confirms that it contains a definition, and ensures that it type checks. Then, for each input, $a_i$, if a corresponding value, $b_i$, is passed, OEIS-LT directly tries to prove a theorem of the form $f(a_i)=b_i$ for the parsed definition $f$ contained in the \texttt{src} argument. To do this, OEIS-LT appeals to \texttt{deriveTheorem} from the \texttt{Sequencelib.Meta} package, which in turn relies on the \texttt{oeis\_tactic} described in the previous section. When a value $b_i$ is not passed for the input, $a_i$, OEIS-LT instead leverages \texttt{deriveTheoremForIndex}, also contained within the \texttt{Sequencelib.Meta} package. This function first attempts to compute $f(a_i)$ by first constructing a \texttt{syntax} for the
function application, elaborating the syntax into an \texttt{Expr}, and then using \texttt{evalExpr} to evaluate the term. Once evaluated, it uses \texttt{deriveTheorem} as in the previous case. In both cases, the theorems are proved within a \texttt{withoutModifyingEnv do} context to ensure memory consumption doesn't grow. 

In addition to direct socket calls, OEIS-LT provides a Python software development kit (SDK) to simplify interactions from within a Python program. The SDK provides native Python functions and types for each of the primary OEIS-LT commands. The following listing illustrates examples usage of the SDK.

\begin{lstlisting}[caption={Using the OEIS-LT Python SDK}]
from oeislt import Client
  
src = "def PowersOfTwo (n: Nat)..."
values = [(0, 1], (1, 2), ...]

# create an OEIS-LT client object
c = Client()

# call eval command
rsp = c.eval(src=src, values=values)

# check the response 
print(rsp.status)
\end{lstlisting}

Some of the initial OEIS-LT commands, particularly the \texttt{gen} command, were designed in part to support a computational pipeline to autoformalize OEIS sequence definitions. This pipeline builds on the work of \cite{gauthier2023learning} which used machine learning to generate thousands of OEIS sequence definition in a small domain-specific language based on Standard ML. The OEIS-LT \texttt{gen} command supports translating function definitions from the domain-specific language defined in \cite{gauthier2023learning} to Lean. 
We describe these aspects in more detail in Section \ref{subsec:pipeline-details}.

The OEIS-LT architecture is highly extensible---new functionality can be added simply by implementing additional commands---and several additional commands are either planned or already in development. These are described in more detail in Section \ref{sec:future}. To define a new command, all that is required is to define a Lean function, $f: \alpha \to \text{OEISM}\,\beta$, where $\alpha$ and $\beta$ are types that implement the \texttt{ToJson} and \texttt{FromJson} type classes, respectively. The monad \texttt{OEISM} is provided by the OEIS-LT package, and it includes error handling and \texttt{IO} lifting . Moreover, OEIS-LT supports a ``plugin'' architecture where an individual instance of the server can be configured with different commands to support different workflows and use cases. To this end, OEIS-LT organizes commands into ``core" commands and ``extension'' commands, whereby ``extension'' commands, and their implementation functions, can be configured to be sourced from different code repositories. 

\subsection{OEIS-LT Performance}
\label{subsec:perf}
We evaluated the performance of an instance of the OEIS-LT server on 
an Ubuntu VM equipped with an Intel Haswell 2.5GHz 16 core processor and 32 GB of RAM. We used the \texttt{locust} framework to simulate concurrent load from virtual users, where each ``user" executed commands against the running OEIS-LT instance in a separate thread. The results from the empirical evaluation are shown in Table~\ref{fig:performance}. For each value $n=8, 25, 50, 100, 200, 300$, we started $n$ virtual user client threads and evaluated a specific OEIS-LT command for a sustained period of 30 seconds. Commands evaluated included \texttt{compile}, \texttt{eval}, and \texttt{prove}. During execution, locust measured the median latency (in miliseconds) and throughput (in requests per second) while a separate \texttt{top} process measured CPU and memory consumption of the OEIS-LT process. Across all executions of all commands, OEIS-LT achieved 100\% successful response rate---zero requests resulted in failures. 

The data in Table~\ref{fig:performance} establish the strong performance of OEIS-LT under load and its ability to scale to additional users with minimal resource utilization. In particular, OEIS-LT achieved a sustained 116 requests per second for its most complex command, \texttt{prove}, with under 2.5 cores of average CPU utilization (a little under 5 cores peak). Additionally, the evaluation shows that OEIS-LT maintains a relatively low memory footprint of around 5-6 GB for most of the tests. All performance results can be reproduced using the scripts available within the evaluation repository \cite{sequencelib_eval_github}.

\begin{table}[t]
\caption{Performance of OEIS-LT commands under various loads.}
\label{fig:performance}
\centering
\begin{tabular}{rrrrrrrl}
\toprule
\textbf{\begin{tabular}{@{}c@{}}Client\\ Threads\end{tabular}} & 
\textbf{\begin{tabular}{@{}c@{}}Median\\ Latency\\ (ms)\end{tabular}} & 
\textbf{\begin{tabular}{@{}c@{}}Median\\ Throughput\\ (req/s)\end{tabular}} & 
\textbf{\begin{tabular}{@{}c@{}}Max\\ CPU\\ (\%)\end{tabular}} & 
\textbf{\begin{tabular}{@{}c@{}}Avg\\ CPU\\ (\%)\end{tabular}} & 
\textbf{\begin{tabular}{@{}c@{}}Max\\ MEM\\ (Gb)\end{tabular}} & 
\textbf{\begin{tabular}{@{}c@{}}Avg\\ MEM\\ (Gb)\end{tabular}} & 
\textbf{\begin{tabular}{@{}c@{}}Command \end{tabular}}\\ 
\midrule
8 & 10 & 6.25 & 70 & 9.43 & 5.25 & 5.24 & compile \\ 
25 & 11 & 19.51 & 110 & 21.61 & 5.38 & 5.35 & compile \\
50 & 11 & 37.68 & 155 & 58.91 & 5.48 & 5.46 & compile \\
100 & 12 & 68.47 & 230 & 104.18 & 5.77 & 5.68 & compile \\
200 & 19 & 108.56 & 360 & 179.36 & 6.32 & 6.04 & compile \\
300 & 21 & 119.36 & 410 & 181.97 & 6.78 & 6.45 & compile \\ 
\midrule
8 & 12 & 6.51 & 110 & 13.76 & 5.25 & 5.25 & eval    \\
25 & 12 & 20.17 & 70 & 24.52 & 5.41 & 5.4 & eval    \\
50 & 13 & 37.46 & 136 & 61.55 & 5.51 & 5.5 & eval    \\
100 & 16 & 68.76 & 250 & 130.91 & 5.77 & 5.77 & eval    \\
200 & 23 & 108.68 & 410 & 193.94 & 6.45 & 6.41 & eval    \\
300 & 28 & 119.84 & 420 & 200.36 & 7.01 & 6.91 & eval    \\ 
\midrule
8 & 15 & 6.48 & 46 & 9.67 & 5.31 & 5.25 & prove   \\
25 & 16 & 19.32 & 109 & 36 & 5.41 & 5.39 & prove   \\
50 & 16 & 38.03 & 210 & 97.27 & 5.51 & 5.5 & prove   \\
100 & 21 & 68.18 & 310 & 165.58 & 5.77 & 5.76 & prove   \\
200 & 38 & 106.67 & 470 & 247.61 & 6.42 & 6.4 & prove   \\
300 & 48 & 116.13 & 490 & 241.58 & 7.01 & 6.97 & prove  \\ 
\bottomrule
\end{tabular}
\end{table}

\subsection{A Computational Pipeline to Autoformalize 25,000 OEIS Sequences}
\label{subsec:pipeline-details}

In \cite{gauthier2023learning}, the authors develop a self-learning loop that iteratively generates, tests and trains a tree neural network to produce function definitions whose values coincide with randomly selected sequences from the OEIS. The loop synthesizes function definitions in a domain-specific language consisting only of basic arithmetic operators, operators representing programming constructs such as \texttt{loop} and \texttt{compr} (comprehension), variables \texttt{x} and \texttt{y}, and the constants $0, 1,$ and $2$. A full scale run of the self-learning loop generated function definitions for 27,987 sequences from the OEIS. An example function generated for the sequence A011000 is given in listing 7. The source code for the self-learning loop and the generated functions is available in the repository \cite{oeis_synthesis_github}.

\begin{lstlisting}[caption={The Standard ML function generated for A011000}]
loop(
  \(x,y).(((2 * loop(\(x,y).x * y, 2 + 2, x)) - x) div y) + x, x, 1)
\end{lstlisting}

We developed a computational pipeline to translate the function definitions from the domain-specific language to Lean, evaluate the function against known values for the corresponding sequence from the OEIS, and try to prove theorems about the values. These primary steps correspond to the OEIS-LT commands \texttt{gen}, \texttt{eval}, and \texttt{prove}, respectively, described previously. Thus, the basic architecture of our pipeline is given in the following listing:

\begin{lstlisting}[caption={Psudeo-code for the Sequencelib pipeline}]
from oeislt import Client 
    
sequences = read_sml_solutions()
c = Client()
    
for s in sequences: 
    src = c.gen(s)           
    r = c.eval(src, values) 
    if r.success:
        thms = c.prove(s)
        write_lean_file(src, thrms)
\end{lstlisting}

The pipeline begins by generating the corresponding Lean code for a sequence definition from the domain-specific language using the \texttt{gen} command from OEIS-LT. 
The \texttt{gen} command makes use of a parser, defined directly in Lean as a set of inductive types corresponding to the constants, variables, and operators in the domain-specific language, together with a new Lean syntax category, \texttt{oeis\_synthesis}, which describes the allowable syntax. It also makes use of functions (e.g., \texttt{DSLToLean}, \texttt{DSLToLeanSimplified}) that elaborate the syntax into Lean expressions. These objects are available from \texttt{Grammar.lean} within the OEIS-LT repository.

Once a Lean function is generated for a given sequence, the pipeline uses the \texttt{eval} command from OEIS-LT to evaluate the function against all values in the main OEIS entry for the sequence. Furthermore, if a b-file exists for the sequence, the pipeline samples a uniform distribution of values from the b-file and evaluates the function on those. A uniform distribution of 100 values was chosen to reduce the computational complexity required to evaluate functions on the entire set of values from the b-files, which could be very large. If any of the values did not agree with the expected value from the OEIS entry, the function was discarded.

Finally, if all values of the Lean function agreed with the known values from the OEIS for the sequence, the pipeline called the OEIS-LT \texttt{prove} command to try and prove theorems about the values. To avoid timeouts due to computational limits, the pipeline first tried to prove 100 values (or the maximum length of known values for the sequence), then 50, then 25. In many cases, the 100 theorems proved could have easily been expanded. This will be done in future work. 

In total, our pipeline formalized 25,457 sequences into Lean definitions, all but 2,520 of the original 27,987 standard ML functions from the original pipeline. Of the 2,540 sequences that were not translated, the vast majority were sequences whose values did not agree with the OEIS values. The pipeline also skipped sequences that were defined in the OEIS using a negative offset. This was due to a small technical limitation that did not impact many sequences (65 in total) and will be addressed in future work. Some additional statistics are given in the following table.

\begin{table}[t]
\caption{Summary statistics for the Lean pipeline.}
\label{table:pipeline_summary}
\centering
\begin{tabularx}{.95\textwidth}{@{}X@{}}
\toprule
\# Sequences from Original Paper~\cite{oeis_synthesis_github} \dotfill 27,987 \\ 
\# Sequences Formalized into Lean \dotfill 25,457 \\ 
\# Sequences with Max Theorems Proved \dotfill 24,546\\  
\# Sequences with Some (Less than Max) Theorems Proved \dotfill 908 \\
\# Sequences with Values Disagreeing (skipped) \dotfill 2,462 \\
\# Sequences Timed Out Checking Values (skipped) \dotfill 3 \\
\# Sequences with Negative Offset (skipped) \dotfill 65 \\
\bottomrule
\end{tabularx}
\end{table}

\section{Future Directions for SequenceLib}
\label{sec:future}
Our goal for Sequencelib---formalizing all of the sequences in the OEIS---presents a tremendous challenge. Many OEIS sequences contain some of the oldest open problems in mathematics, such as the Goldbach Conjecture (A002372), the Collatz Conjecture (A087003), the Busy Beaver numbers (A028444), and many more. Thus, we fully expect the development of Sequencelib to be the focus of future work for decades. Our near-term plans for future development broadly fall into three categories: 1) AI-enabled pipelines for \textit{autoformalization}, i.e., the task of generating a formalization of a mathematical statement and proof provided in natural language; 2) Expansion of the \texttt{oeis-tactic} and other metaprogramming tools to support \textit{proof synthesis}, i.e., the task of constructing a proof of a formal mathematical statement; 3) Software engineering efforts to ensure the Sequencelib platform reamins scalable. We describe each of these areas briefly in what follows. 

\textit{Autoformalization.} Multiple additional AI-based pipelines leveraging different methods are in development with the goal of making additional contributions to Sequencelib. First, we are developing a pipeline based on frontier models to autoformalize the definitions of sequences from their natural language definition. Using additional metadata available from the OEIS entry and examples from Sequencelib together with realtime feedback from OEIS-LT, preliminary results suggest that frontier models may be able to autoformalize the definition of a large portion of the OEIS---perhaps as many as 100,000 sequences or more. As with the pipeline described in the previous section, the new pipeline will evaluate a Lean function candidate against the known OEIS values for the sequence to ensure correctness. Second, we plan to enhance the frontier pipeline using a joint embedding model that aligns natural language and formal language pairs in a common semantic space to guide the formalization of additional sequences. 

\textit{Metaprogramming for Proof Synthesis.} Enhancements to the \texttt{oeis\_tactic} will support deriving theorems about the values of sequences. For example, we plan to incorporate additional tactics from mathlib, such as \texttt{ring\_nf}, to simplify expressions and enable additional proof synthesis cases to be solved. We also plan to expand the tactic to support synthesizing proofs that two functions representing the same OEIS sequence coincide. Such a technique will be especially useful as new AI pipelines autoformalize sequences that have been previously formalized, but with potentially different function definitions. 

We are also studying a graph derived from the OEIS, where the nodes of the graph are given by sequences, and an undirected edge from two sequences, $a_n$ and $b_n$, constitutes a theorem relating the values $a_n$ to the values of $b_n$. Some preliminary results suggest this graph contains a small, strongly connected component. These notions and metadata as well as additional tactics and other metaprogramming tools will be bundled into a new release of OEIS-LT tool server. Furthermore, we will be releasing a benchmark dataset based on Sequencelib that can be used to evaluate AI pipelines, ITPs, and tool servers, such as OEIS-LT.

\textit{Software Engineering for the Sequencelib Platform.} The Sequencelib platform consists of much more than just the Lean files comprising the library and includes the build system, test suite, website, documentation site, OEIS-LT tool server, etc. In order to scale to tens of thousands of sequences, as generated by the initial ML pipeline described previously, several engineering efforts were required. As the library grows to another order of magnitude, we fully anticipate needing additional engineering effort to ensure the platform scales accordingly.

\section{Conclusion}
This paper presented the Sequencelib project, a software library, computational 
toolchain, and web-based platform to formalize the mathematics contained within 
the On-Line Encyclopedia of Integer Sequences (OEIS) in the Lean interactive 
theorem prover. Formalization of the OEIS provides significant benefits but also
presents a formidable challenge. To help address this challenge, we developed and 
presented OEIS-LT, a new tool server specialized for the task of formalizing 
OEIS sequences. Furthermore, er described a computational pipeline that leveraged 
OEIS-LT to formalize more than 25,000 sequence and prove move than 1.6 million 
theorems about their values. As future work, we are developing additional computational pipelines based on frontier AI models and OEIS-LT that we expect to lead to the formalization of thousands of additional OEIS sequences. 
\label{sec:conclusion}

\bibliography{references}

\end{document}